\documentclass[12pt,prb,superscriptaddress,amsmath,amssymb,floatfix]{revtex4} 
\usepackage{graphicx}
\usepackage{xspace}

\newcommand{\nick}{Ni\ensuremath{_{147}}\xspace}
\newcommand{\doped}{C-doped Ni\ensuremath{_{147}}\xspace}
\newcommand{\tot}{$\langle\,$E$_{tot}\,$$\rangle$\xspace}
\newcommand{\tfr}{T$_{freezing}$\xspace}
\newcommand{\tml}{T$_{melting}$\xspace}
\newcommand{\frtml}{T$_{freezing}$$\,<\;$T$\;<\;$T$_{melting}$\xspace}
\newcommand{\intpot}{\ensuremath{V^{Ni-C}(r)}\xspace} 

\begin{document}
\title{Impurity effects on the melting of Ni clusters}

\author{Andrey Lyalin}
\altaffiliation[On leave from  ]
{Institute of Physics, St. Petersburg State University, Ulianovskaya Str. 1, 198504 St. Petersburg, Petrodvorez, Russia}
\email[Email address: ]{lyalin@fias.uni-frankfurt.de}
\affiliation{Frankfurt Institute for Advanced Studies, Goethe-University, Ruth-Moufang-Str. 1, 60438 Frankfurt am Main, Germany}
\affiliation{Imperial College London, Physics Department, Prince Consort Road, London SW7 2BW, UK}

\author {Adilah Hussien}
\affiliation{Frankfurt Institute for Advanced Studies, Goethe-University, Ruth-Moufang-Str. 1, 60438 Frankfurt am Main, Germany}

\author {Andrey V. Solov'yov}
\altaffiliation[On leave from  ]
{A. F. Ioffe Physical-Technical Institute, Polytechnicheskaya 26, 194021 St. Petersburg, Russia}
\affiliation{Frankfurt Institute for Advanced Studies, Goethe-University, Ruth-Moufang-Str. 1, 60438 Frankfurt am Main, Germany}

\author {Walter Greiner}
\affiliation{Frankfurt Institute for Advanced Studies, Goethe-University, Ruth-Moufang-Str. 1, 60438 Frankfurt am Main, Germany}

\date{\today}

\begin{abstract}
We demonstrate that the addition of a single carbon impurity 
leads to significant changes in the thermodynamic properties 
of Ni clusters consisting of more than a hundred atoms. 
The magnitude of the change induced is dependent upon the 
parameters of the Ni--C interaction. Hence, thermodynamic 
properties of Ni clusters can be effectively tuned by the 
addition of an impurity of a particular type. We also show 
that the presence of a carbon impurity considerably changes 
the mobility and diffusion of atoms in the Ni cluster at temperatures 
close to its melting point. The calculated diffusion coefficients of the 
carbon impurity in the Ni cluster can be used for a reliable estimate of 
the growth rate of carbon nanotubes.
\end{abstract}


\maketitle

\section{Introduction}
\label{intro}

Thermodynamic properties of atomic clusters depend considerably on the cluster size. 
It has been shown that the melting temperature of a small spherical particle
decreases with the reduction of its 
radius.\cite{Thomson71,Pawlow1909,Takagi54,Hanszen60,Qi01}
This is due to the substantial increase in the relative number of weakly bounded atoms on the surface 
in comparison with those in the bulk. Such a size effect on the melting temperature of small
particles of different metals having diameters down to 2-3 nm 
has been confirmed in series of experiments.\cite{Buffat76,Castro90,Lai96,Bottani90} 

However for clusters having sizes smaller than 1-2 nm,
the melting temperature is no longer a monotonic function of the cluster size. 
Experiments on sodium clusters Na$_N$, with number of atoms $N=50-360$,
have demonstrated that melting temperature as a function of size shows a prominent 
irregular structure with local 
maxima.\cite{Schmidt97,Schmidt98,Kusche99,Haberland05} 
The origin of the non-monotonic variation in the melting temperature with respect to cluster size lies 
in interplay between electronic and geometric shell effects in the sodium 
clusters.\cite{Haberland05}  
Intensive theoretical efforts have been undertaken to identify the details of the geometric 
and electronic structures underlying the variations in the melting 
temperature.\cite{Calvo00,ReyesNava03,Calvo04,Manninen04,Chacko05,Aguado05a,Aguado05b,Noya07}

Experiments on small cluster ions of tin \cite{Shvartsburg00} and gallium \cite{Breaux03} have 
confirmed the violation of the linear relationship between the reduction in the melting temperature 
and the inverse radius of the cluster. It was discovered that the melting temperature of selected Sn$_N$ and Ga$_N$ clusters, of sizes $N<40$, can 
considerably exceed the melting temperature of the corresponding bulk 
material.\cite{Shvartsburg00,Breaux03} 
This behavior was explained by the structural differences between the small clusters 
and the bulk.\cite{Shvartsburg00}

Modification of a cluster structure can also
be induced by the doping of an impurity or substitution of one or 
several atoms in a pure cluster with an atom of a different type.\cite{Darby02} 
It has been shown that doping of sodium clusters 
with Li, Cs and O$_2$ impurities (as well as Al$_{13}$ and  Ga$_{13}$ clusters with a carbon impurity) 
results in a decrease
of the melting temperature of the cluster.\cite{Aguado04,Chandrachud07,Hock08} 
Whereas doping of the icosahedral silver clusters with Ni and Cu atoms considerably increases 
the melting temperature.\cite{Mottet05}
Therefore one can suppose that the thermodynamic properties 
of a selected cluster can be tuned by doping with an impurity of a particular type.

In this paper we report the results of a systematic theoretical study 
regarding the effect of impurity on 
the thermodynamic properties of Ni clusters.
We demonstrate that adding a single carbon impurity can result in 
changes in the melting temperature of an \nick cluster. The magnitude of the change induced is dependent upon the 
parameters of the interaction between the Ni atoms and the C impurity. 
Hence, thermodynamic properties of Ni clusters can be effectively 
tuned by the addition of an impurity. 
We also show that the presence 
of a carbon impurity considerably changes the mobility of atoms in 
the Ni cluster at temperatures close to its melting point. 

The choice of Ni clusters is stipulated by their high 
chemical and catalytic reactivity, unique properties 
and multiple applications in nanostructured materials.\cite{Goddard07}
An important example of such an application is the process of 
the catalytically activated growth of carbon nanotubes. 
The mechanism of this process is not yet well understood 
(see, e.g., Refs. \onlinecite{Golovko05,Obolensky07a,Obolensky07} and references therein)
and knowledge of the specific role of the impurity in the Ni
catalytic nanoparticle may ascertain whether the carbon nanotube structure 
and its growth kinetics can be controlled.
The thermodynamic state of the catalytic nanoparticle plays a crucial 
role on the carbon nanotube growth.\cite{Harutyunyan05} 
The nanotube growth rate can be obtained by a solution of a set of kinetic 
equations which include, in particular, the 
diffusion flux of carbon through the metal particle.\cite{Louchev03,Baker78,Obolensky07} 
On the other hand the diffusion coefficients depend 
on the thermodynamic state of the catalytic Ni cluster which can, in turn, be 
tuned by doping with an impurity. 
Therefore direct molecular dynamics study of carbon diffusion in the Ni cluster is 
important for a reliable estimation 
of the growth rate of carbon nanotubes.

\section{Theoretical methods}
\label{theory}

The study of structural and dynamical properties of
the clusters of transition metals is a challenging task due to 
the presence of unfilled valence \emph{d}-orbitals. 
The high density of the \emph{d}-states and their localized character make 
the direct \emph{ab initio} methods computationally very demanding for 
clusters larger than several dozens of atoms.\cite{Nayak97} 
In order to describe  the structure of clusters of larger sizes,
one needs to use approximate methods and model interatomic potentials.

An effective approach for study of transition-metal clusters is the 
embedded-atom method \cite{Daw83,Daw84,Finnis84,Foiles86,Sutton90,Sutton96} 
which takes into account many-body effects. 
The latter appears through the inhomogeneous electron density of the system.  
In this paper, the molecular dynamics study of the \nick cluster 
has been performed using the Sutton-Chen \cite{Sutton90} many-body potential 
which belongs to the family of the embedded-atom types of potentials. 
The Sutton-Chen potential has been shown to reproduce bulk and surface 
properties of transition metals and their alloys with sufficient 
accuracy (see, e.g., Refs. \onlinecite{Sutton91,Todd93,Lynden95,Nayak97,Doye98} 
and references therein). 
The potential energy of the finite system within
the Sutton-Chen model has the following form:

\begin{equation}
U_{pot}= \varepsilon \sum_i \left[ 
\frac{1}{2} \sum_{j \ne i} \left( \frac{a}{r_{ij}} \right)^{n} - c \rho_{i}^{1/2}  
\right] ,
\end{equation}

\noindent where

\begin{equation}
\rho_{i} =  \sum_{j \ne i} \left( \frac{a}{r_{ij}} \right)^{m} .
\end{equation}

\noindent Here $r_{ij}$ is the distance between atoms $i$ and $j$, 
$\varepsilon$ is a parameter with dimension of energy,
$a$ is the lattice constant, $c$ is a dimensionless parameter, and $n$ and $m$ are positive
integers with $n > m$. The  parameters provided by Sutton and Chen
for nickel have the following values:\cite{Sutton90}
$\varepsilon = 1.5707 \cdot 10^{-2}$ eV,
$a = 3.52$ \AA\,
$c = 39.432$,
$n=9$, and $m=6$.

The determination of a reliable model potential for the Ni -- C interaction 
is a difficult task. 
In Refs. \onlinecite{Yamaguchi99,Shibuta07,Limia07} the binding energy and different 
charge state of various forms of $NiC_n$ clusters (n= 1 - 3)  were determined
using \emph{ab initio} density functional calculations based on the three-parameter Becke-type gradient-corrected
exchange functional with the gradient-corrected correlation
functional of Lee, Yang, and Parr (B3LYP) \cite{Becke_PRA88,Becke_JCP93,LYP_PRB88}
and LANL2DZ basis set (see, e.g., Ref. \onlinecite{Chemistry} and references therein). This was done 
in order to develop a many-body potential for the transition metal -- carbon interaction.
The many-body potential was then successfully used for a molecular dynamics study of the formation 
of metallofullerenes \cite{Yamaguchi99} and single-walled carbon nanotubes.\cite{Shibuta07} 

In the present work, however, the interaction between Ni atoms and C impurity is modeled by the 
Morse pair potential:
\begin{equation}
V^{Ni-C} (r) = \varepsilon_{M} \left( \left(1-e^{\rho (1-r/r_0) }\right)^2 - 1  \right).
\label{Morse}
\end{equation}

\noindent We have determined the parameters of the Morse potential 
($\varepsilon_{M}= 2.431 $ eV, $\rho= 3.295$, $r_0=1.763$ \AA\ ) by fitting
the Ni -- C interaction obtained in Refs. \onlinecite{Yamaguchi99,Shibuta07,Limia07}
within the B3LYP/LANL2DZ method. The choice of using the
pairwise Morse potential is due to its simplicity which allows us to study the influence of 
the parameters in (\ref{Morse}) on the thermodynamic properties
of the \doped clusters, keeping a clear physical picture of the process occuring in the system.

The optimized initial geometries of the cluster have been determined by 
finding local minima on the multidimensional
potential energy surface. We have applied an efficient scheme 
of global optimization, called the Cluster Fusion Algorithm 
(CFA). \cite{LJ_PRL03,LJ_IJMPE04,CoLe05}
The CFA belongs to the class of genetic  
global optimization methods, \cite{Goldberg89,Michalewicz96} which are 
very promising for structural optimization of nanoalloys, see, e.g., Ref. \onlinecite{Ferrando08} 
and references therein. 
The scheme has been designed within the context 
of determination of the most stable cluster
geometries and it is applicable for various 
types of clusters. \cite{LJ_PRL03,struct_Mg,magnetism_La,struct_Sr}

Molecular dynamics simulations have been performed for the canonical (NVT) 
ensemble of particles: 
the number of particles N, the volume V and the temperature T 
of the system are kept constant.
Integration of Newton's equations of motion have been done using the Verlet leapfrog algorithm,
with a time step $\Delta t=$ 1~fs and a total simulation run of 10~ns (excluding an initial equilibration time of 50~ps).
For temperature control, we have used the Nos\'e-Hoover 
thermostat \cite{thermostat:nose,thermostat:hoover} 
because it correctly generates a canonical ensemble.

\section{Numerical results and discussion}
\label{results}

\begin{figure}[htb]
\includegraphics[scale=0.8]{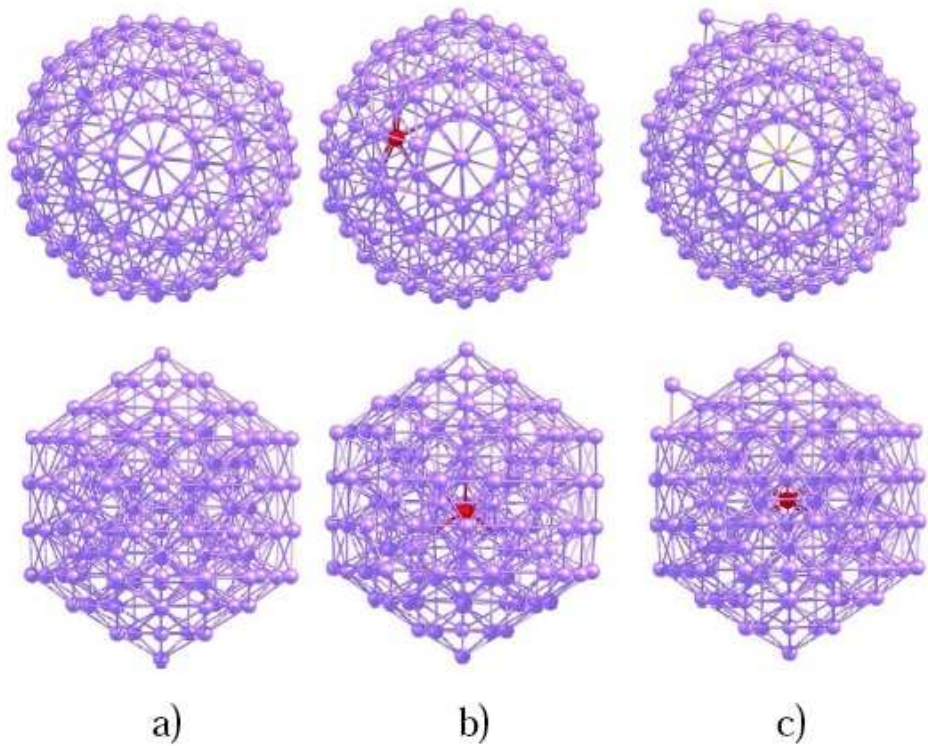}
\caption{(Color online) --- (a) Optimized structure of a pure \nick cluster;  
(b) the \emph{isomer} state structure of the \doped cluster;  
(c) the \emph{ground} state structure of the \doped cluster.}
\label{fig:structures}
\end{figure}

During the last, decade numerous theoretical and experimental efforts
have been devoted to studying structural, electronic, and magnetic 
properties of Ni clusters (recent comprehensive reviews can be 
found in Refs. \onlinecite{Baletto_RMP05,Alonso05}). 
It has been demonstrated  that for small Ni clusters,
a motif based on the icosahedral 
structure dominates the cluster growth, at least around sizes of complete
Mackay icosahedra.\cite{Pellarin94,Lopez96,Doye98,Zhang05}
The results of the geometry optimization for the pure and the \doped clusters
are shown in Fig. \ref{fig:structures}. 
The ground state of the \nick cluster is a perfect icosahedron, Fig.~\ref{fig:structures}(a), in agreement with the results of 
the previous works (see Refs. \onlinecite{Alonso05,Zhang05} and references therein).

For the \doped cluster we found that 
the lowest energy state possesses the icosahedral type of structure with
the C atom in the central position and one Ni atom outside the filled icosahedral shell,
as shown in Fig.~\ref{fig:structures}(c). 
The energetically closest isomer of the \doped cluster 
is the deformed icosahedron with the C impurity located in the vicinity of the cluster center,
as depicted in Fig.~\ref{fig:structures}(b).
The impurity creates a local distortion between the first and the second closed 
icosahedral shells. 
The difference in the binding energy per atom 
between the ground, Fig.~\ref{fig:structures}(c), and the isomer, Fig.~\ref{fig:structures}(b), states of the \doped 
cluster is 0.002 eV. 
The isomer state, Fig.~\ref{fig:structures}(b), is metastable because 
in order to transform the cluster structure 
from the isomer, Fig.~\ref{fig:structures}(b), to its ground state, Fig.~\ref{fig:structures}(c),
it is necessary to overcome the energy barrier.
The isomer, Fig.~\ref{fig:structures}(b), can be naturally formed by successive migration of the outer C 
atom from the cluster surface towards its center. 
Namely, this type of isomer are presumably formed in carbon nanotube growth experiments
where a molecule of a feedstock gas or a free C atom collides with the surface of the Ni particle. 
Therefore, we have performed molecular dynamics simulations for the initial  
geometry corresponding to the isomer state, Fig.~\ref{fig:structures}(b) of the \doped cluster.

\subsection{Pure \nick clusters}

\begin{figure}[htbp]
\includegraphics[scale=1.2]{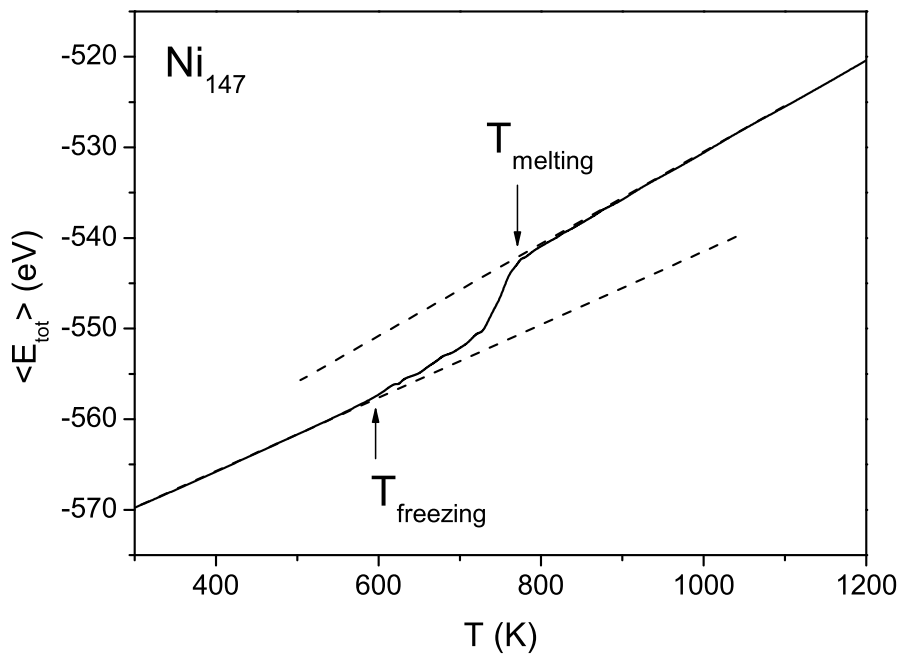}
\caption{Caloric curve for the pure \nick cluster. Temperatures T$\,<\,$\tfr and 
T$\,>\,$\tml correspond to the completely frozen and melted states, respectively.}
\label{fig:caloric_Ni147}
\end{figure}

\begin{figure}[htbp]
\includegraphics[scale=1.2]{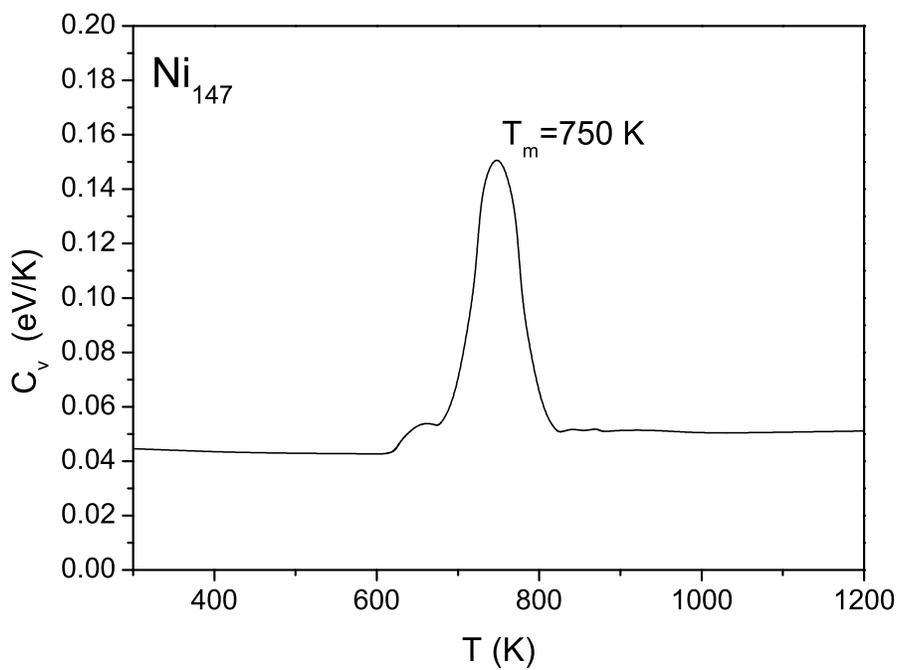}
\caption{Heat capacity for the pure \nick cluster as a function of $T$. 
Temperature $T_m = 750$~K indicates the melting temperature of the 
cluster (defined as the maximum of the heat capacity). }
\label{fig:heat_Ni147}
\end{figure}

Figures~\ref{fig:caloric_Ni147} and \ref{fig:heat_Ni147} 
demonstrate the temperature dependence of the time-averaged total energy \tot (i.e. the caloric curve) and the 
heat capacity $C_{V}$ calculated for the \nick cluster, respectively. 
The heat capacity at constant volume 
is defined as a derivative  of the internal energy over the temperature:
$C_{V} = \left( \partial E / \partial T \right)_V$.

A thermal phase transition is indicated in the caloric curve by a change in
the gradient of the  temperature-dependent total energy \tot. 
The height of the jump near the phase transition point gives an estimate of the 
latent heat, \emph{i.e.} the energy which is associated with 
the destruction of the ordered lattice. 
Figure~\ref{fig:caloric_Ni147} demonstrates that the change
in the slope of the caloric curve takes place in the wide interval of temperatures
between \tfr $\approx$ 600~K and \tml $\approx$ 800~K. 
For the temperatures  T$\,<\,$\tfr, the cluster is completely frozen;
while for temperatures T$\,>\,$\tml, the cluster has melted.  
The intermediate interval \frtml corresponds to the mixed 
state where the solid and the liquid phases coexist. This is a typical behavior 
of the caloric curve for finite systems, in contrast to the abrupt stepwise jump
of the caloric curve corresponding to the melting temperature of a bulk.    

The melting process can also be easily 
recognized by the peak in the temperature dependence of the 
heat capacity, as seen in Fig.~\ref{fig:heat_Ni147}. 
We have found that for the pure \nick cluster 
the melting temperature associated with the maximum of 
the heat capacity $T_m=750$~K is considerably smaller than the melting 
temperature for the bulk nickel $T_m^{bulk}=1728$~K.\cite{meltbulkNi}  
As was discussed above, the decrease in the melting temperature of the 
finite size clusters in comparison with the bulk occurs due to 
a substantial increase in the relative number of weakly bounded 
atoms on the cluster surface.
According to the so-called 'Pawlow law' the 
melting temperature of a spherical particle possessing a homogeneous surface
decreases linearly with increasing its inverse radius:\cite{Pawlow1909}
\begin{equation}
T_{m}=T_{m}^{bulk} \left(  1 - \frac{\alpha}{R} \right),
\label{eq:pawlow}
\end{equation}
\noindent where $R$ is the radius of the spherical particle and $\alpha$ is a constant,
which can be defined by fitting the temperature $T_{m}$ to available experimental data. 
Equation (\ref{eq:pawlow}) is valid for the relatively
large particles when shell effects are not important.
The similar reduction in the melting temperature 
of the large Ni$_N$ clusters with the number of atoms $N$ ranging from 
336 to 8007 has been reported in Ref. \onlinecite{Qi01}.
Based on the calculated melting temperatures from Ref. \onlinecite{Qi01}, we have estimated
the values of $T_{m}^{bulk}$ and $\alpha$ to be 1590 K and 3.65 \AA{} respectively. 
The radius of the cluster has been defined as $R=r_{s} N^{1/3}$, where $r_s=1.375$ \AA{} is 
the Wigner-Seitz radius for bulk nickel. \cite{Kittel}

Apart from the main peak, the heat capacity displays an additional
maximum at $T \approx 660$~K---suggesting a stepwise melting process. This additional maximum  
corresponds to a slight change in the slope of the caloric 
curve beginning at \tfr $\approx$ 600~K. Such a situation corresponds 
to the so-called \emph{pre-melted} state---when the cluster surface melts first while 
the core of the cluster remains frozen. 
{ This is the stationary state where the coexistence of two phases---liquid surface 
and frozen cluster core---is observed. The exact delimitation 
of the two phases is relative but can be defined from the 
difference in mobility of atoms taken from 
the cluster surface and its core.}
Visualization of the molecular dynamics trajectories of atoms
located at the surface and in the core of the cluster confirms this interpretation.

\begin{figure}[hp]
\includegraphics[scale=0.8]{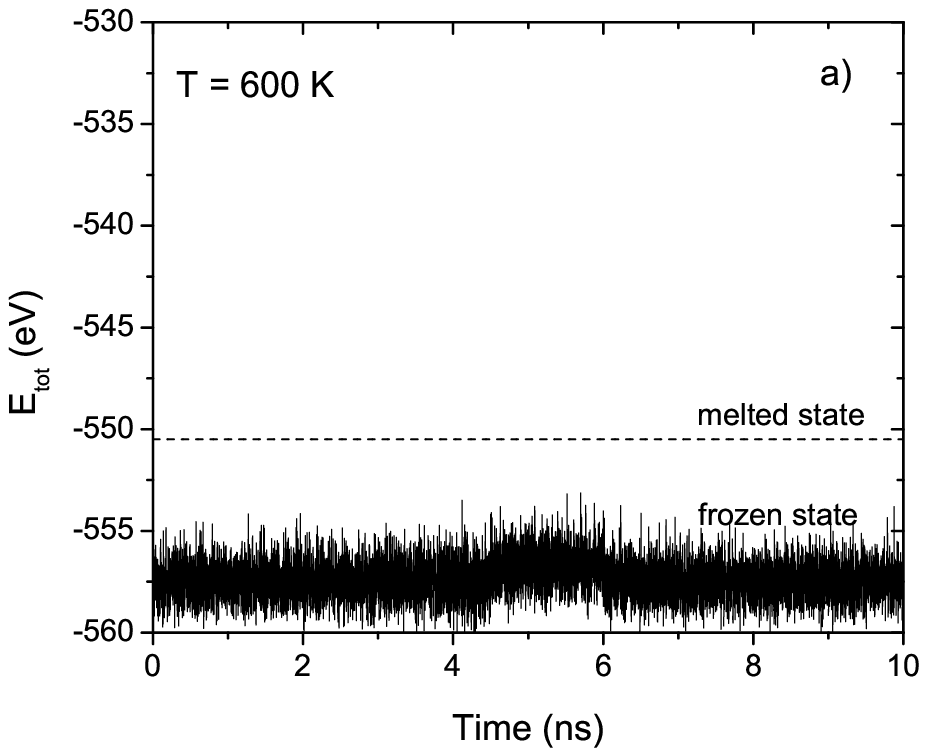}
\includegraphics[scale=0.8]{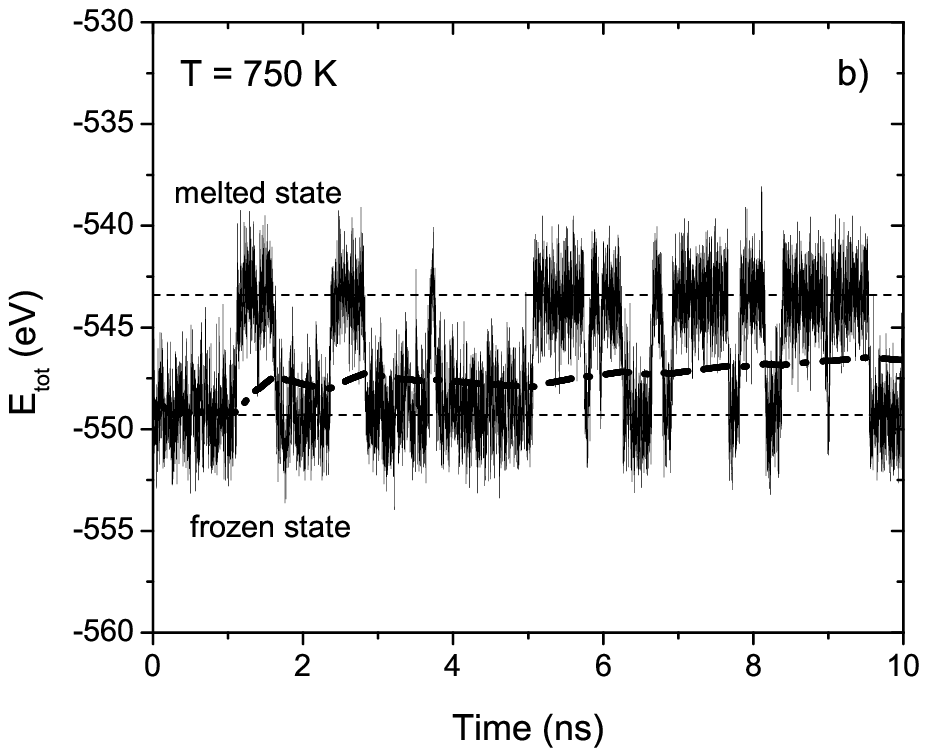}
\includegraphics[scale=0.8]{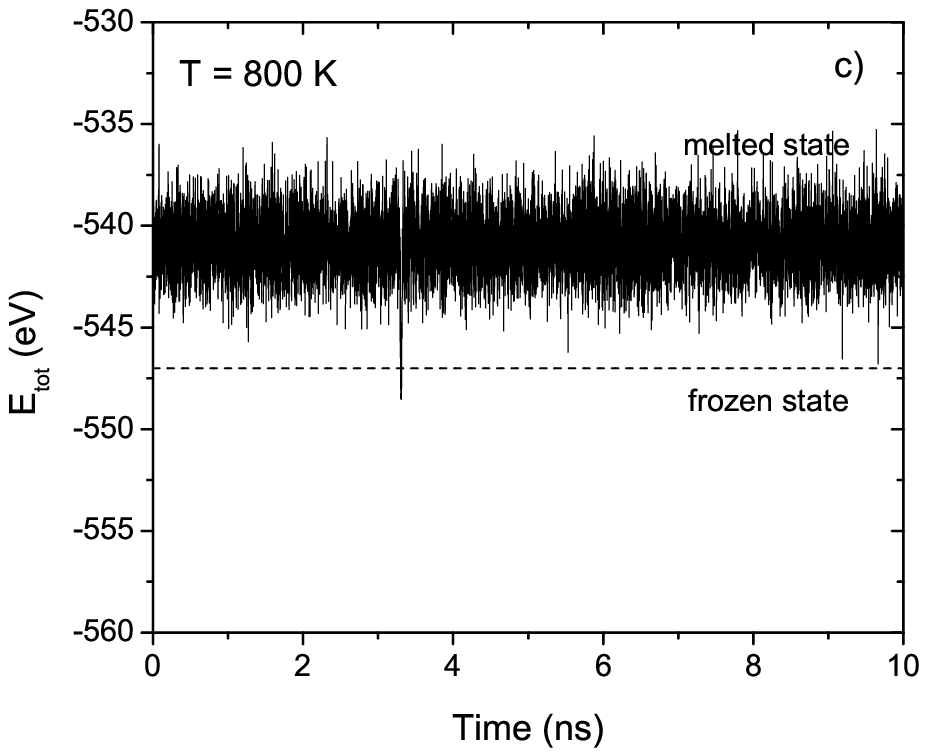}
\caption{Time dependence of the instantaneous values of the total energy calculated for the \nick cluster: 
(a) T$\:=600$~K, frozen state;
(b) T$\:=750$~K, phase transition;
(c) T$\:=800$~K, molten state.
The dashed-dotted lines demonstrate the time-averaged total energy of the system.}
\label{fig:fluctuations}
\end{figure}

Figures~\ref{fig:fluctuations}(a), \ref{fig:fluctuations}(b), and \ref{fig:fluctuations}(c) 
present the time dependence of the 
instantaneous values of the total energy 
calculated for the \nick cluster at 
the temperatures 600~K, 750~K, and 800~K, 
respectively. The total simulation time is 10~ns.

It is seen from Fig.~\ref{fig:fluctuations}(a) that  
the instantaneous values of the total energy oscillate around their 
time-averaged value at the cluster temperature $T= 600$ K. 
The fluctuations in the total energy of the system 
is a result of choosing the canonical ensemble of particles. 
To reproduce this ensemble, the temperature of the system 
is controlled by the Nos\'e-Hoover thermostat, 
which produces the appropriate 
energy fluctuations.\cite{thermostat:nose,thermostat:hoover} 
The frozen phase is characterized 
by thermal vibrations of the atoms around equilibrium 
positions in the cluster lattice, while the overall 
cluster structure remains unchanged.

For the temperature $T= 750$~K, corresponding to the maximum 
of the heat capacity, the time dependence of the instantaneous values
of the total energy changes, 
as shown in Fig.~\ref{fig:fluctuations}(b).  
The total energy of the system initially oscillates around its typical
averaged value for the frozen \nick state. 
After some time, the system
jumps to a disordered molten state---causing the 
total energy of the system to oscillate around
the average value typical for the molten \nick state.
The system then, after a finite period in the molten state,
jumps back to its initial frozen state. This behavior is repeated and the system 
continously oscillated between the frozen-molten states. 
Note that, in the interval of temperatures \frtml,
it is necessary to perform the molecular dynamics simulations
for a relatively long time (on the order of 10 ns) to achieve time independence 
of the averaged total energy \tot of the system.
The average life-time of the system in 
the frozen and the molten  
states depends upon the temperature of the cluster, 
resulting in a smooth change of the \tot from the 
frozen to the molten state as a function of temperature.
Thus, Fig.~\ref{fig:fluctuations}(b) clearly demonstrates coexistance of 
the two thermodynamic phases of the finite system at the temperature 
of phase transition.
A similar behavior of the time dependence of the short-time potential energy averages
calculated for the $Ar_{13}$ cluster at the temperature 33~K has been reported in 
Refs. \onlinecite{Jellinek86,Jellinek_Ar13}. 

Finally, for the temperature $T= 800$~K and above,
the total energy of the \nick cluster 
oscillates around its typical averaged value for the molten
state.

\subsection{\doped clusters}

We now focus our study on the thermodynamic properties 
of the \doped cluster. 
We consider a cluster with the C impurity in the vicinity of the cluster center
as shown in Fig.~\ref{fig:structures}(b).

\begin{figure}[htbp]
\includegraphics[scale=1.2]{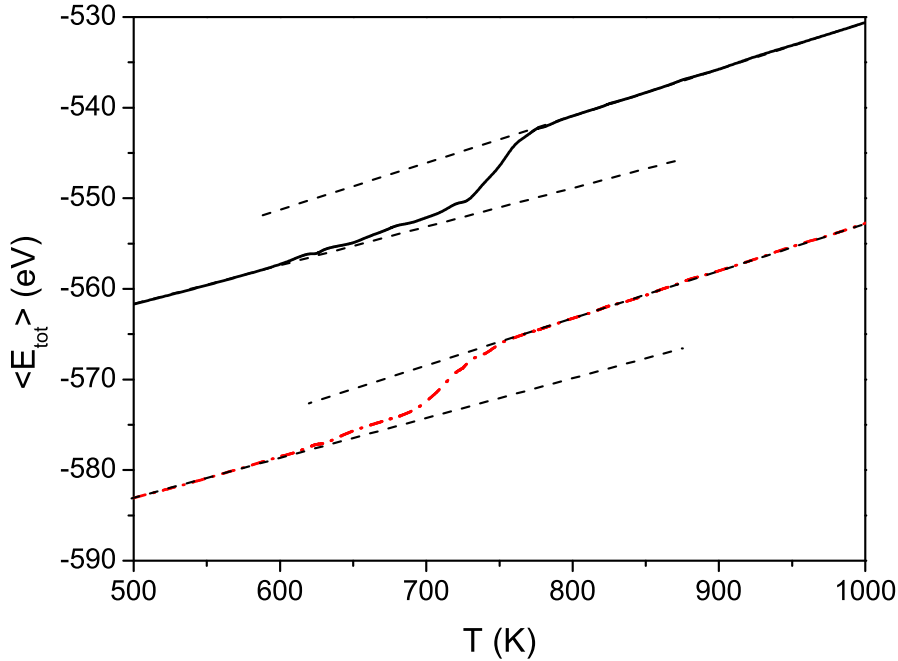}
\caption{Caloric curve for the pure \nick cluster (solid line) 
and the \doped cluster (dashed-dotted line).}
\label{fig:caloric_Ni147_C}
\end{figure}

\begin{figure}[htbp]
\includegraphics[scale=1.2]{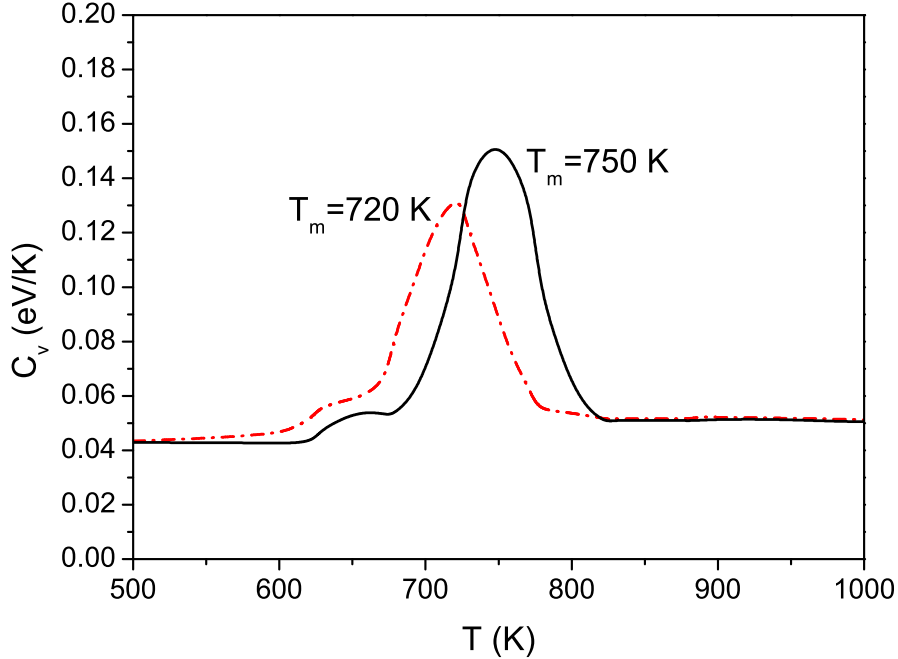}
\caption{Heat capacity for the pure \nick cluster (solid line) 
and the \doped cluster (dashed-dotted line) as a function of $T$. 
Temperatures $T_m = 750$~K and $T_m = 720 $ K indicate the melting temperatures 
of the pure and the \doped cluster, respectively.}
\label{fig:heat_Ni147_C}
\end{figure}

Dashed-dotted lines in Figs. \ref{fig:caloric_Ni147_C} and \ref{fig:heat_Ni147_C} 
demonstrate the temperature dependence of the 
caloric curve and the heat capacity calculated 
for the \doped cluster. 
For the sake of comparison, we have plotted the same dependencies calculated for the pure 
\nick cluster in Figs.~\ref{fig:caloric_Ni147_C} and \ref{fig:heat_Ni147_C} by solid lines. 
As seen in Figs.~\ref{fig:caloric_Ni147_C} and \ref{fig:heat_Ni147_C},
doping of the \nick cluster with a single C impurity reduces its melting temperature by 30~K. 
Thus, doping of a cluster consisting of more than a hundred of atoms by just one 
additional atom of impurity results in a considerable change of its melting temperature.

Figure~\ref{fig:caloric_Ni147_C} demonstrates that the 
first change in the slope of 
the temperature-dependent \tot calculated for 
the \doped cluster occurs at a temperature 
slightly below 600~K. This fact indicates
that the C impurity does not have much influence on the surface melting of the cluster. 
Indeed, the C impurity is 
located near the cluster center where it induces local deformations. However, it does not significantly change
the structure of the cluster surface, see Fig.~\ref{fig:structures}(b); 
therefore, it does not have much of an effect on surface melting. 

Further increase in the cluster temperature results in the melting of the \doped
cluster with the melting temperature $T_m = 730$ K. The latter is 
defined by the position of the maximum of the temperature-dependent heat capacity. 
The decrease in the melting temperature of 
the \doped cluster can be explained as a result of the local 
distortions of the cluster's icosahedral structure near the impurity. 
The C atom introduces 
a strain in the cluster lattice which decreases stability of the cluster and thus, 
its melting temperature. 

This effect is in accordance with the behavior of bulk material where it is known that defects in 
crystalline lattice provide nucleation sites for the liquid state 
(see, e.g., Ref. \onlinecite{Ferrando08} and references therein).
However such an effect strongly depends upon the type of impurity. 
As was demonstrated in Ref. \onlinecite{Mottet05}, doping of the Ag$_{55}$ cluster with Cu 
and Ni impurities in the center 
(where the bond length between Ag atoms and atoms of impurity are considerably 
smaller than the bond length between Ag atoms \cite{Ferrando08,Mottet05})
results in an increase of the melting temperature of the system. 
Therefore substitution of the central Ag atom with 
an atom of Cu or Ni impurity  induces the
contraction of the cluster lattice in the vicinity of the center, which can 
partially release the strain of the icosahedral structure.\cite{Mottet05} 
Thus the small impurity in the center allows the cluster 
to relax toward a configuration with better 
interatomic distances, thereby increasing the cluster stability.\cite{Ferrando08,Mottet05}

Our results demonstrate that doping of the C atom in the vicinity 
of the center of the \nick cluster induces an 
additional strain in the cluster structure and, as a result, the melting temperature 
of the cluster decreases.
To study in detail how this induced strain can change the thermodynamic behavior 
of the \doped cluster, we have performed calculations
of the melting temperature of the cluster 
for the set of different parameters $\rho$ and $r_0$ in the Morse potential (\ref{Morse})---in other words,
modeling the variation in the Ni -- C interaction.

\begin{figure}[htbp]
\includegraphics[scale=1.2,clip]{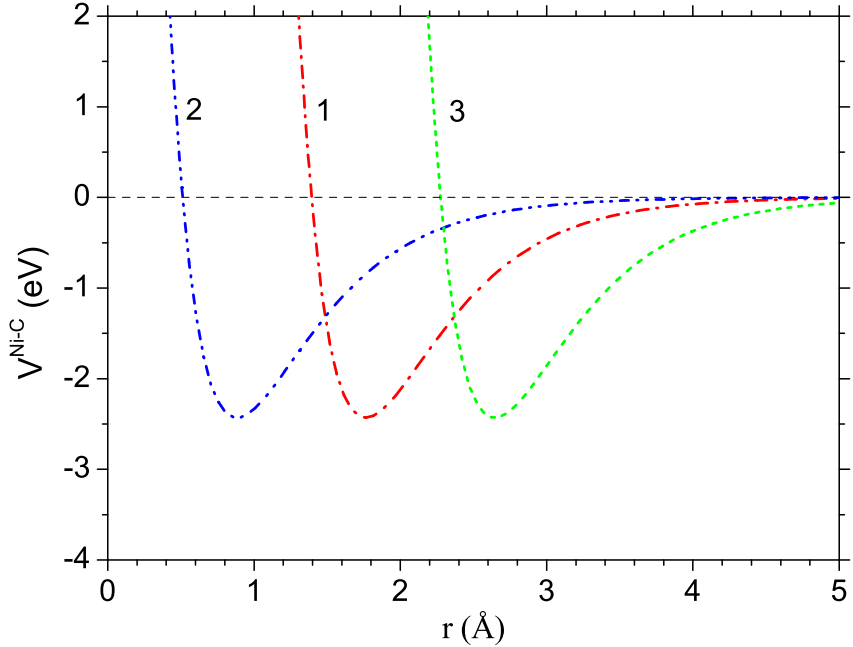}
\caption{(Color online) Morse potential for the Ni -- C interaction with different values of parameters 
$\rho$ and $r_0$. \emph{Curve 1}, $\rho= 3.295$, $r_0=1.763$ \AA\ (optimal values); \emph{Curve 2}, $\rho= 1.648$, $r_0=0.882$ \AA\ (reduced bonding); 
\emph{Curve 3}, $\rho= 4.943$, $r_0=2.645$ \AA\ (enlarged bonding). The depth of the potential well is kept constant $\varepsilon_{M}= 2.431 $ eV.}
\label{fig:Morse_var}
\end{figure}

Curve 1 in Fig.~\ref{fig:Morse_var} presents the dependence of the Morse potential 
\intpot on the interatomic distance $r$ caclulated for the optimal values of the parameters  
$\varepsilon_{M}$, $\rho$, and $r_0$.
Curve 2 in Fig.~\ref{fig:Morse_var} presents the potential \intpot when the 
bond constant $r_0$ and parameter $\rho$ are reduced by the factor 0.5 
in comparison to their optimal values; 
while curve 3 in Fig.~\ref{fig:Morse_var} presents the interaction potential \intpot
when the parameters $r_0$ and $\rho$ are enlarged by the factor 1.5.
Hence, the potentials presented in Fig.~\ref{fig:Morse_var} 
by curves 2 and 3 model the impurity effect with the local compression and 
expansion of the cluster lattice respectively.


\begin{figure}[htb]
\includegraphics[scale=0.8]{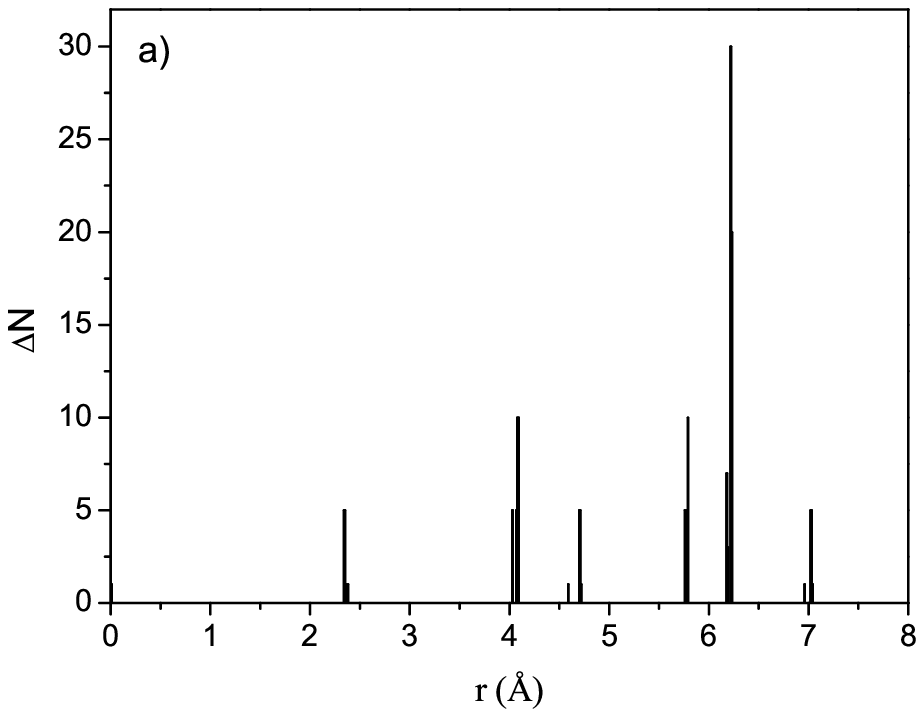}
\includegraphics[scale=0.8]{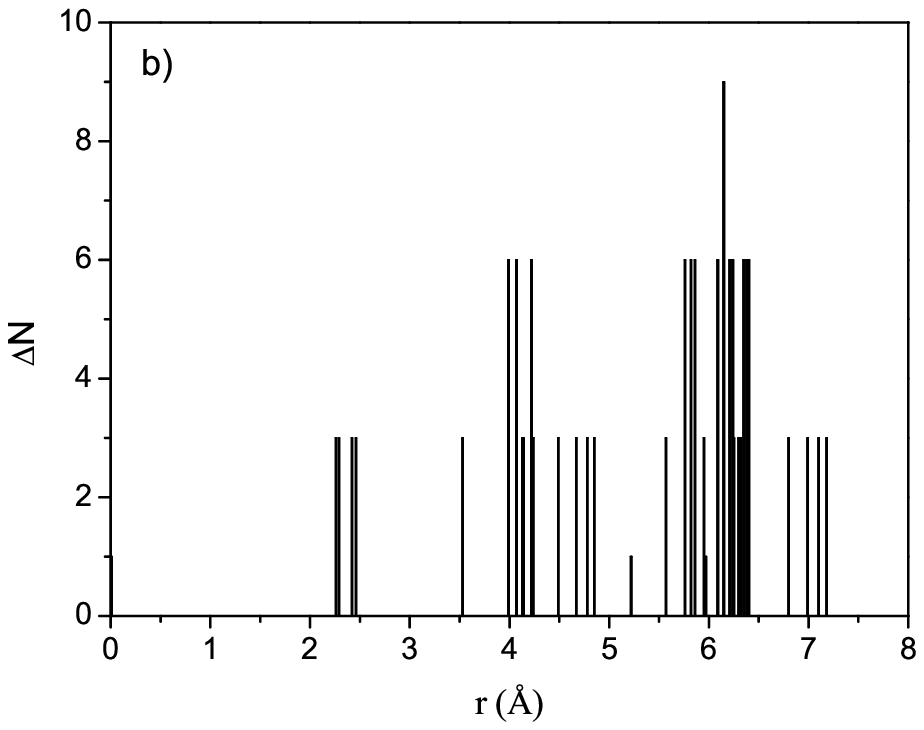}
\includegraphics[scale=0.8]{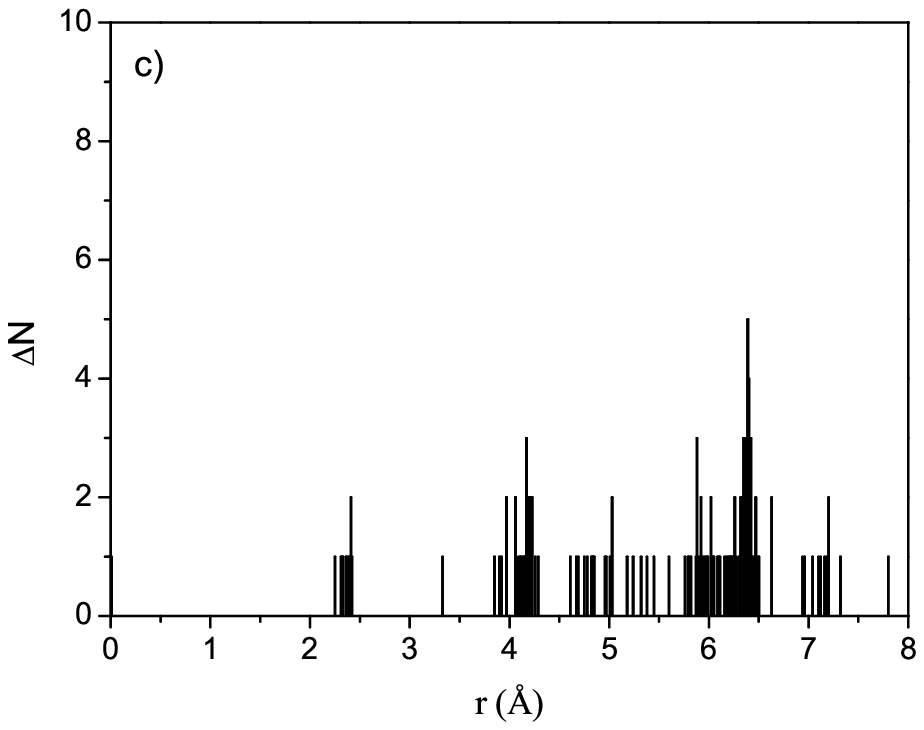}
\includegraphics[scale=0.8]{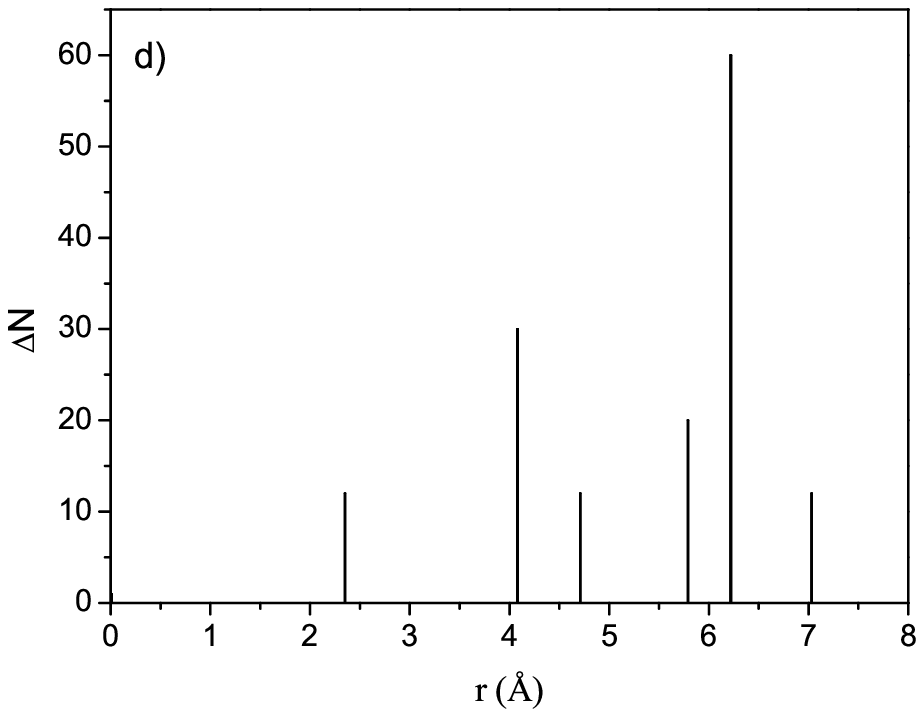}
\caption{The number of Ni atoms
$\Delta N$ at distances between $r$ and $r + \Delta r$ from the center of mass
of the \doped cluster calculated for different values of the parameters
$\rho$ and $r_0$:
(a) $\rho= 1.648$, $r_0=0.882$ \AA{} (reduced bonding);
(b) $\rho= 3.295$, $r_0=1.763$ \AA{} (optimal values);
(c) $\rho= 4.943$, $r_0=2.645$ \AA{} (enlarged bonding);
{(d) pure \nick cluster for comparison.}
The radial interval $\Delta r$ is 0.001 \AA{}.}
\label{fig:hist}
\end{figure}

To illustrate the effect of variation in the Ni -- C
interaction on cluster structure
we present, in Figs.~\ref{fig:hist}(a)-(c), the histogram 
of the radial distribution of the number of Ni atoms
located at distances between $r$ and $r + \Delta r$ from the center of mass
of the \doped cluster (where the width of the radial interval
is $\Delta r$ = 0.001 \AA{}). The change in the cluster structure, 
due to varying the parameters of
$\rho$ and $r_0$ in the Morse potential, can thus be clearly seen.
{ For comparison, Fig.~\ref{fig:hist}(d)
represents the histogram of the radial distribution of the number of Ni atoms
in the pure \nick cluster. }

In the case of the reduced bonding between the C impurity and the Ni atoms,
the overall icosahedral shell structure of the cluster remains preserved---although
there exists some relaxation of the lattice, 
Fig.~\ref{fig:hist}(a).
The increase of the effective radius of the Morse potential for the
Ni -- C interaction results in strong distortions of the
icosahedral shell structure of the cluster,
Figs.~\ref{fig:hist}(b) and \ref{fig:hist}(c).
Such distortions reduce the stability of the cluster in comparison
to the compact icosahedral structure.


\begin{figure}[htbp]
\includegraphics[scale=1.2,clip]{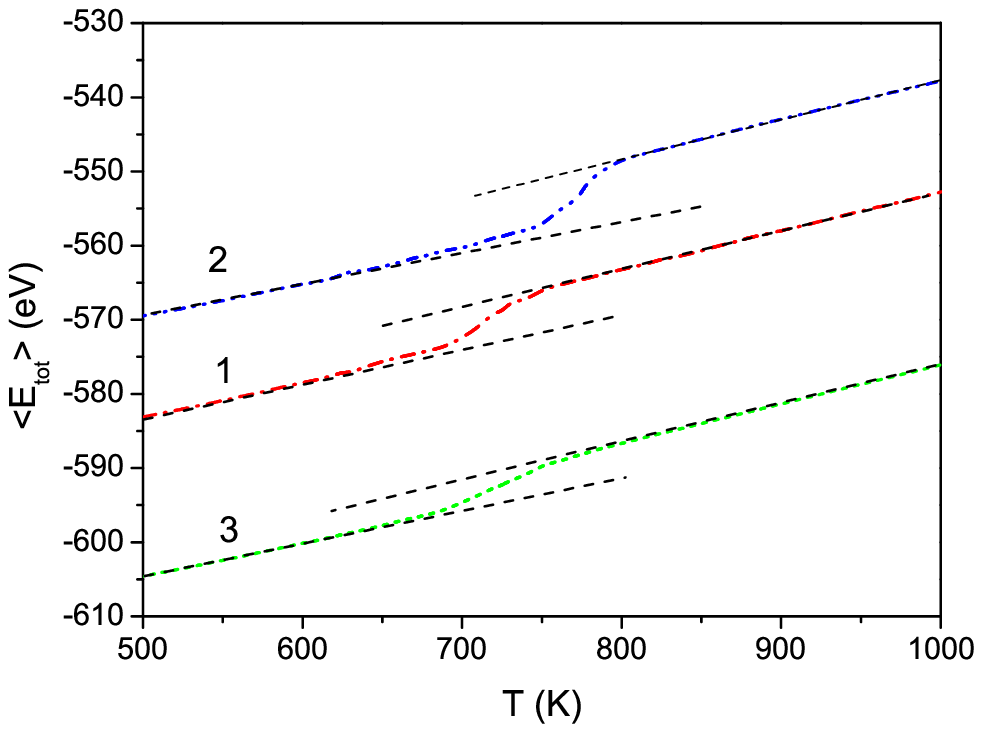}
\caption{(Color online) Caloric curves for the \doped cluster
calculated for the different values of parameters 
$\rho$ and $r_0$.
\emph{Curve 1}, $\rho= 3.295$, $r_0=1.763$ \AA\ (optimal values);
\emph{Curve 2}, $\rho= 1.648$, $r_0=0.882$ \AA\ (reduced bonding); 
\emph{Curve 3}, $\rho= 4.943$, $r_0=2.645$ \AA\ (enlarged bonding). 
The depth of the potential well is kept constant $\varepsilon_{M}= 2.431 $ eV.}
\label{fig:caloric_Ni147_change}
\end{figure}

\begin{figure}[htbp]
\includegraphics[scale=1.2,clip]{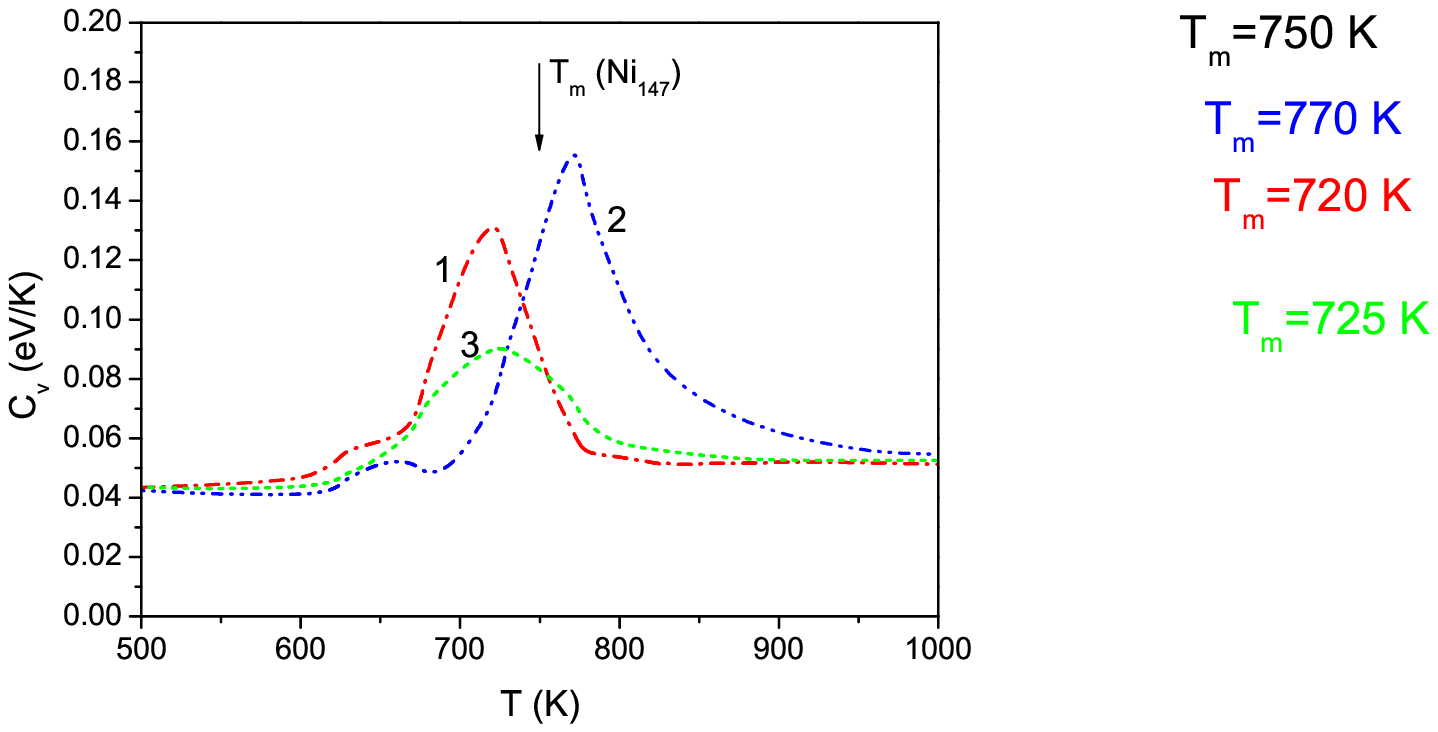}
\caption{(Color online) Heat capacity for the \doped cluster
calculated for the different values of parameters 
$\rho$ and $r_0$.
\emph{Curve 1}, $\rho= 3.295$, $r_0=1.763$ \AA\ (optimal values);
\emph{Curve 2}, $\rho= 1.648$, $r_0=0.882$ \AA\ (reduced bonding); 
\emph{Curve 3}, $\rho= 4.943$, $r_0=2.645$ \AA\ (enlarged bonding). The depth of 
the potential well is kept constant $\varepsilon_{M}= 2.431 $ eV. 
Arrow indicates the melting temperature of the pure \nick cluster.}
\label{fig:heat_Ni147_change}
\end{figure}

Figures~\ref{fig:caloric_Ni147_change} and \ref{fig:heat_Ni147_change} demonstrate 
the temperature dependence of the 
caloric curve and the heat capacity of 
the \doped cluster
calculated for the different parameters 
of the Morse potential \intpot. 
Curves 1 in Figs. \ref{fig:caloric_Ni147_change} and \ref{fig:heat_Ni147_change}
present the caloric curve and the heat capacity 
calculated with the optimal values of parameters  
for the Ni -- C interaction. 

Decreasing the effective radius of the Morse potential (see curve 2, 
Fig.~\ref{fig:Morse_var}) by a factor 0.5 (\emph{c.f.} the optimal values)
results in the contraction of the cluster lattice in the vicinity of impurity.
This contraction releases the strain in the icosahedral structure, therefore 
leading to an increase of the cluster stability. In this case the, melting temperature of the 
impurity doped \nick cluster increases by 21~K, compared the 
pure cluster (see curve 2 in Fig.~\ref{fig:heat_Ni147_change}).

An increase in the effective radius of the Morse potential for Ni -- C interaction 
to its optimal value (see curve 1 in Fig.~\ref{fig:Morse_var})
results in the appearance of an additional strain in the cluster lattice. 
Therefore, the melting temperature of the \doped cluster 
decreases compared to the pure \nick cluster. 

The further increase in the effective radius of the Ni -- C interaction creates 
a strong deformation and rearrangement in the cluster structure,
see Fig.~\ref{fig:hist}(c). In this case, location of the impurity 
in the vicinity of the cluster center becomes energetically unfavorable. Thus, the impurity atom 
shifts towards the cluster surface to minimize the destruction of the icosahedral lattice.  	
This effect prevents the further decrease in the 
melting temperature of the cluster, as an impurity located on the cluster 
surface will not have much influence on the melting transition of the core.


Melting transition can be also recognized from the analysis of 
the trajectories of the individual atoms and their diffusion in the volume of the cluster. 
The melting transition occurs when atoms begin their Brownian motion  
instead of the thermal vibrations 
around their equilibrium  positions in the ordered cluster's lattice. 
Such a transition can be seen as a step in the temperature dependence of the 
diffusion coefficient.
The diffusion coefficient of the atom $i$ in a media is defined as 
(see, e.g., Refs. \onlinecite{Landau6,Jellinek86}):

\begin{equation}
D_{i} = \frac{1}{6} \frac{\rm d}{{\rm d} t } 
\left<  r_i^2(t) \right>,
\label{diffusion_D}
\end{equation}

\noindent where $\left<  r_i^2(t) \right>$ is the mean-square displacement 
averaged along the atom's trajectory

\begin{equation}
\left<  r_i^2(t) \right> = \frac{1}{n_t} \sum_{j=1}^{n_t} 
\left(  
{\bf R}_i(t_{0j}+t) - {\bf R}_i(t_{0j}) 
\right)^2 .
\end{equation}

\noindent Here ${\bf R}_i(t)$ is a radius vector of an atom $i$ at the time $t$, and 
$n_t$ is the number of time origins, $t_{0j}$, considered along the trajectory. 
Equation (\ref{diffusion_D}) is valid for the simulation time smaller 
than the time required for a particle 
to migrate across the diameter of the cluster.

\begin{figure}[hp]
\includegraphics[scale=0.7]{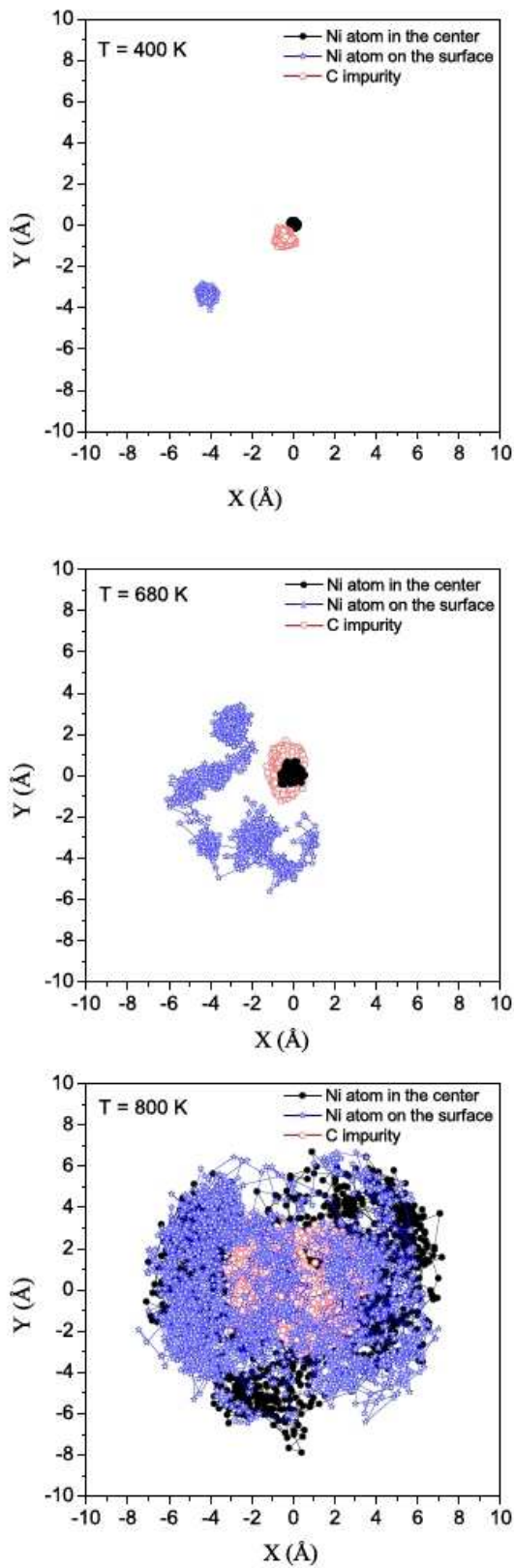}
\caption{(Color online) The 2D projection of trajectories calculated for the Ni atom in the 
center of the \doped cluster (filled dots), the Ni atom { from the vertex}
of the cluster surface  
(stars) and the C impurity (open dots) 
for the cluster temperatures $T=$ 400 K, 680 K, and 800 K. 
The output time step is 1 ps and the
total simulation time is 2 ns.}
\label{fig:trajectory}
\end{figure}

Figure~\ref{fig:trajectory} demonstrates the 2D projection 
of trajectories calculated for 
the Ni atom in the center of the \doped cluster (filled dots), 
the Ni atom { from the vertex} of the cluster surface  (stars) 
and the C impurity (open dots). The temperature of the cluster 
ranges from 400~K to 800~K, as is shown in Fig.~\ref{fig:trajectory}.
The output time step is 1 ps with the total simulation time of 2~ns.

As seen in Fig.~\ref{fig:trajectory}, for a low temperature, T $=400$~K,
all selected atoms in the \doped cluster vibrate 
around their equilibrium positions. At this temperature, 
the cluster is frozen---as per the analysis of the caloric curve and the heat capacity 
calculated for the \doped cluster, see Figs. 
\ref{fig:caloric_Ni147_C} and \ref{fig:heat_Ni147_C}).

It has been discussed above that the temperature dependence of the heat capacity
of the \doped cluster exhibits two maxima corresponding to surface and volume melting of the cluster.
It is seen from Fig.~\ref{fig:heat_Ni147_C} that the first maximum in 
the temperature dependence of the heat capacity
appears when the cluster is at T$=640$~K (surface melting), 
while the second maximum appears at T$=720$~K (volume melting). 
The temperature of 680~K corresponds to the intermediate state when 
the surface of the cluster has already melted but the cluster core is still frozen, as confirmed by 
the analysis of the atomic trajectories. 
As can be seen in Fig.~\ref{fig:trajectory} at T$=680$~K, the 
surface Ni atom begins to diffuse on the surface, 
while the central Ni atom and the C impurity are still vibrating around 
their equilibrium positions. 
{ Note, that the icosahedral surface is inhomogeneous and consists of 
12 vertices, 20 faces and 30 edges. The binding energies of the atoms taken from the 
vertices, faces and edges are thus slightly different. Hence, these atoms begin diffusion 
at different temperatures.}

Finally at the temperature of 800~K, cluster has completely melted as can be observed from the 
temperature behavior of the heat capacity presented in Fig.~\ref{fig:heat_Ni147_C}. 
Figure~\ref{fig:trajectory} demonstrates that the Ni atoms are moving in the entire 
volume of the cluster. The C impurity is also moving randomly in the cluster 
volume although the movement only occurs in 
the central part of the cluster. Hence, this supposes the
heterogeneous distribution of the C atoms in the melted \nick 
clusters at T$=800$~K.  


\begin{figure}[hp]
\includegraphics[scale=0.55]{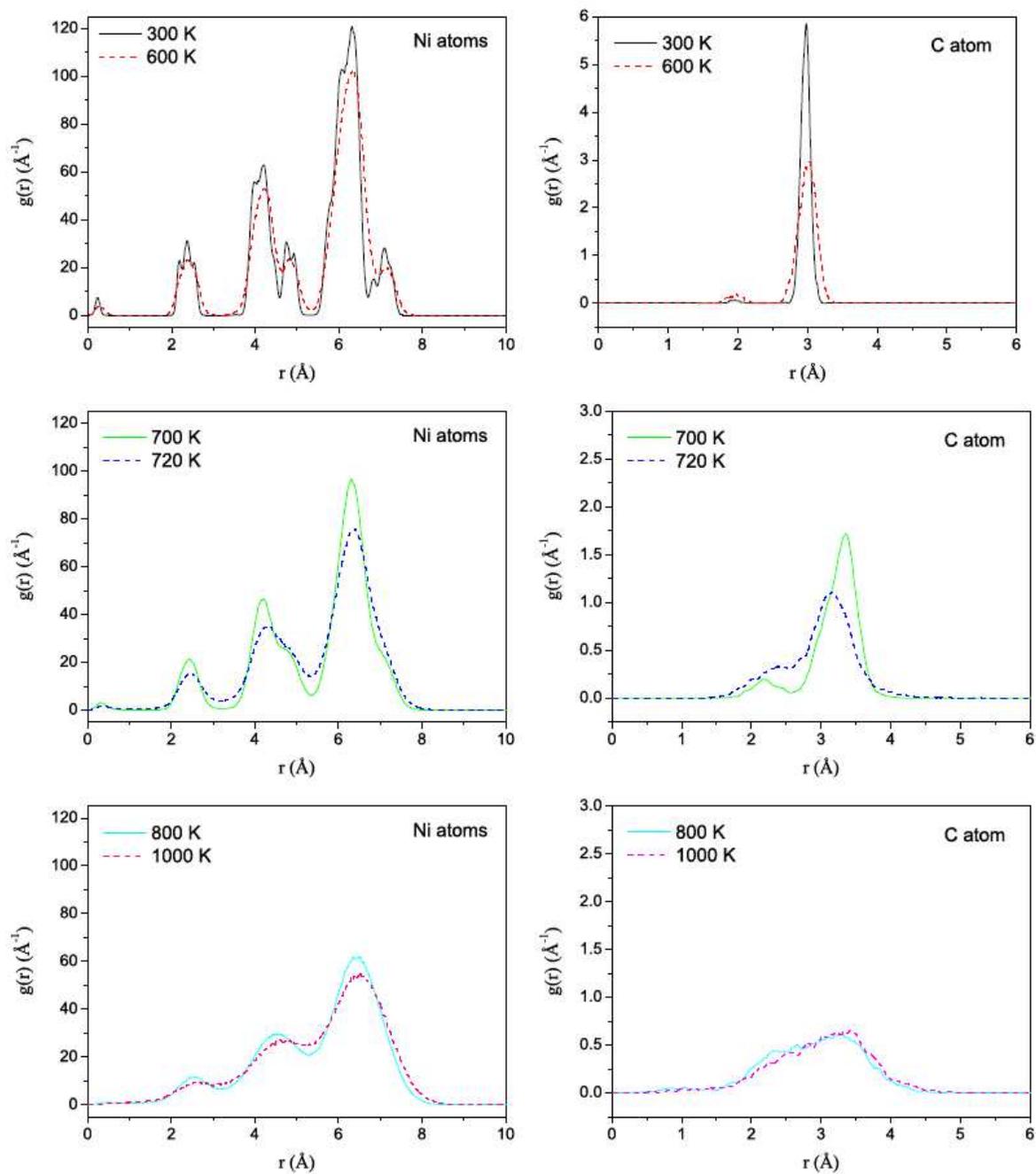}
\caption{ (Color online) The time-averaged
radial distribution function $g(r)$ calculated for the Ni and C atoms in 
the \doped cluster at different temperatures.}
\label{fig:radial_distribution}
\end{figure}

The change in cluster structure upon melting is clearly seen in the time-averaged
radial distribution of atoms in the cluster.
The radial atomic distribution function $g(r)$ is defined as:
\begin{equation}
g(r) = dN(r) / dr,
\end{equation}

\noindent where $dN(r)$ is the number of atoms 
in the spherical layer at distances  
between $r$ and $r + dr$ from the center 
of mass of the cluster.

Figure~\ref{fig:radial_distribution} demonstrates the time-averaged radial distribution function
$g(r)$ for Ni and C atoms in the  \doped cluster calculated at cluster temperatures of 
$T =$ 300~K, 600~K, 700~K, 720~K, 800~K, and 1000~K. The chosen range 
of temperatures allows for the analysis of the cluster structure in the frozen, 
transitional and molten states.

At the low cluster temperature, $T =$ 300~K, one can see the icosahedral shell structure of 
of the nickel subsystem of the \doped cluster, consisting of the central atom 
and three icosahedral shells. The second and third shells are split---corresponding
to atoms in vertex (12 atoms per shell) and non-vertex positions. Heating the cluster   
up to 600~K washes out the subshell splitting, nonetheless the icosahedral shells remain well 
separated. 

With increase in the cluster temperature, up to 700~K, the second and the third shells beging to merge,
although the first and the second shells are still separated (i.e., the radial distribution function 
$g(r)$ is equal to zero in the space between shells). At the temperature corresponding to the maximum in the heat capacity of the \doped cluster, $T=$ 720~K, the first and second icosahedral shells merge. 

Finally, at the temperatures corresponding to the molten state 
(800~K and 1000~K in Fig.~\ref{fig:radial_distribution}), 
the distribution of Ni atoms become more homogeneous and 
the sharp shell structure washes out.  Nevertheless, even at $T=$ 1000~K, 
some radial order with maxima at 2.6 \AA\,, 4.6 \AA\,, and 6.5 \AA\, still remains, 
suggesting that even a molten cluster of a finite size manifests some signs of a 
shell structure.  This effect might be a general feature of finite systems, similar
to the surface-induced ordering in liquid crystals or the layering effect at free 
liquid surfaces.\cite{Aguado04,Ocko86,Chacon01}

Figure~\ref{fig:radial_distribution} demonstrates that, at cluster temperatures 300~K and 600~K, 
the time-averaged radial distribution function calculated for the C impurity 
atom exhibits a sharp maximum at distances 3 \AA\, from the center of mass of the 
\doped cluster. Thus, at these 
temperatures, the C impurity is located between the first and the second icosahedral shells 
of the Ni atoms.  By further increasing the cluster temperature to temperatures near the phase transition 
region, the radial distribution $g(r)$
of the C impurity becomes wider and the appearance of a second maximum 
at distances $\approx$ 2 \AA\, can be observed. While at $T >$ 800~K,
the C impurity can be found to be distributed in the central part of the cluster.


\begin{figure}[ht]
\includegraphics[scale=1.2]{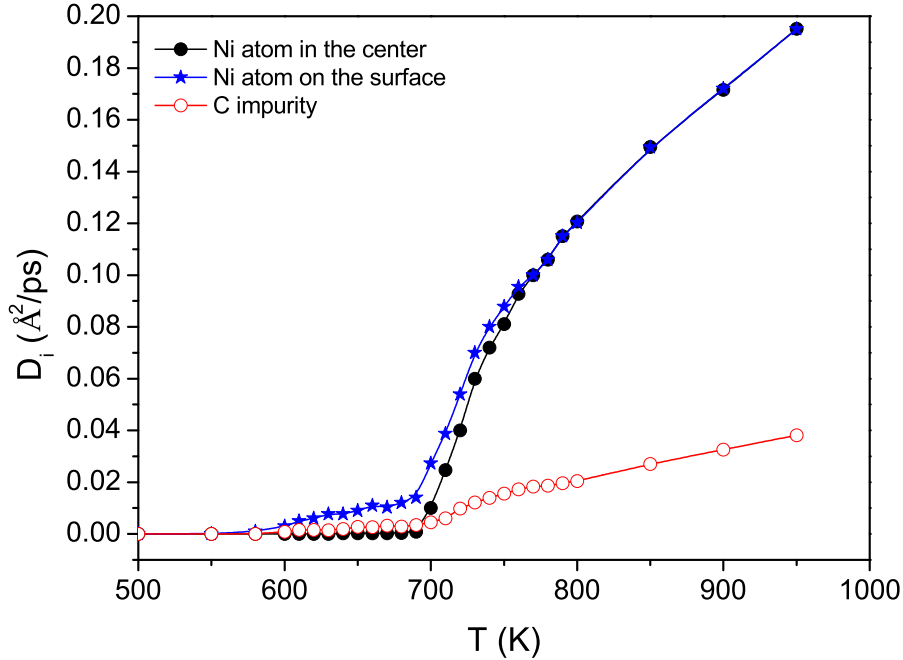}
\caption{(Color online) Temperature dependence of diffusion coefficients calculated 
for the Ni atom in the center of the \doped cluster (filled dots), 
the Ni atom on the cluster surface (stars) 
and the C impurity (open dots).}
\label{fig:diffusion}
\end{figure}

Figure~\ref{fig:diffusion} demonstrates the temperature dependence of 
diffusion coefficients calculated for the selected atoms 
in the \doped cluster.
It is seen from Fig.~\ref{fig:diffusion} that the Ni atom on the cluster 
surface begins to diffuse at temperature of 600~K, 
while Ni atom in the cluster center and the C impurity remain frozen. 

As has been shown above, the caloric curve calculated for the \doped cluster 
(see Fig.~\ref{fig:caloric_Ni147_C}) clearly demonstrates a
stepwise melting behavior. 
In the temperature interval $600\,<\,$~T~$\,<\,700$~K, the slope of 
the caloric curve changes slightly in comparison with that for the frozen state, 
suggesting the existence of a pre-melted state. For the temperatures 
$700\,<\,$~T~$\,<\,760$  the time-averaged total energy \tot growing rapidly
and finally reaches its asymptotic behavior at T$\;>760$ K. The variations in the slope
of the caloric curve result in the appearance of the two maxima in the temperature dependence 
of the heat capacity. These maxima are associated with the surface and core (volume)
melting of the \doped cluster. This proposition is fully confirmed by the
analysis of the temperature dependence of the 
diffusion coefficients for the Ni atoms from: a) the cluster surface and b) the cluster center.

Thus, in the temperature interval of $600\,<\,$~T~$\,<\,700$~K, only surface atoms 
diffuse, confirming that the initial melting of the cluster surface. The Ni atom located in the 
cluster center then begins to diffuse at T$=700$~K. The diffusion coefficients 
calculated for Ni atoms located on the cluster surface and in the core are different 
up to $T=760$~K. 
The difference in the diffusion coefficients show that the surface and the core atoms 
are not yet fully mixed in the cluster volume.  
The difference disappears at cluster temperatures of $T>760$~K, 
when the \doped cluster has become completely molten.  

Figure~\ref{fig:diffusion} demonstrates that the C impurity begins to diffuse 
at T$\;\approx 700$~K---similar to the central Ni atom. 
However, values of the diffusion coefficients calculated for the C impurity are considerably
lower than those for the Ni atoms. 
Knowledge of the diffusion coefficients of the C impurity in a nickel cluster 
can be used for building a reliable kinetic model of carbon nanotube 
growth.\cite{Louchev03,Obolensky07}

\section{Conclusion}
\label{conclusion}

Doping of \nick with a carbon impurity lowers its 
melting temperature  by 30 K due to excessive stress on the cluster lattice. 
The magnitude of the change induced is dependent upon the 
parameters of the interaction between the nickel atoms and the carbon impurity. 
We have demonstrated that an induced contraction of the icosahedral cluster's
lattice in the vicinity of the impurity results in an increase of the melting temperature 
of the cluster;
whereas additional strain in the lattice results in 
the reduction of the melting temperature.  
Therefore, the melting temperature of atomic clusters can be effectively 
tuned by the addition of an impurity of a particular type. 

Doping by a C impurity changes the melting temperature of the cluster,
consequently this means that doping affects the mobility of the atoms in the Ni cluster.
This effect has to be taken into consideration in particular applications 
with metal clusters when the entire process depends on the thermodynamic state 
of the cluster. An example of such experiment is 
the process of the catalytically activated growth of carbon nanotubes.
The kinetics of the carbon nanotube growth depends upon 
diffusion of carbon atoms through the metal catalyst.
Presence of the impurities can considerably change the flux, thereby affecting 
the growth rate of the carbon nanotube. 
The additional change in the thermodynamic state of 
the catalytic particle in the nanotube growth process 
might also depend on the strength of the 
interaction of the particle with a substrate.

In the present work, we have considered a single C impurity 
in the cluster of \nick. We intend to further study the effect of how 
several C impurities will influence the thermodynamic properties 
of the host cluster. In particular, it is 
important to find the optimum conditions (concentration of C atoms, temperature, 
thermodynamic state of the particle, etc.) when the C atoms begin 
aggregating into ordered carbon structures, such as nanotubes.

The influence of impurities on properties of finite systems is a general effect.
While our results were obtained for free clusters, many interesting problems can be found
when one considers the influence of impurities on the phase transitions 
and stability of clusters deposited on a substrate. 
Thus, recently it has been experimentally shown
that the oxidation of silver clusters deposited on a
HOPG surface changes the stability and morphology of cluster 
formations.\cite{Lando06,Lando07} 
Clusters on substrates have important technological applications and the
understanding of how these clusters stabilize 
on the surface are of profound interest.

\begin{acknowledgments}
This work is partially supported by the European Commission
within the PECU project (contract No. 4916 (NEST)) and 
Network of Excellence project EXCELL.
The authors gratefully acknowledge support by the 
Frankfurt Center for Scientific Computing. 
\end{acknowledgments}

\end{document}